%% file: main.tex
\documentclass[a4paper,11pt]{article}

\usepackage{graphicx}
\usepackage{subcaption}
\usepackage{float}
\usepackage{adjustbox}  
\usepackage{textcomp}
\usepackage{booktabs} 
\usepackage{array}    
\usepackage{caption}  
\usepackage{siunitx}  

\usepackage{graphicx} 

\usepackage{float} 
\usepackage{tabularx}
\usepackage{siunitx}
\usepackage{subcaption}

\usepackage{comment} 

\usepackage{graphicx}
\usepackage{subcaption}

 \usepackage{float}  

\usepackage{xcolor}
\usepackage[normalem]{ulem}

\usepackage{jinstpub} 
\usepackage{lineno}

\title{Design and Evaluation of a PMT High-Voltage system for Deepsea Neutrino Telescope}

\author[d]{Zhu~Mao,\footnote[1]{\label{co_author}These authors contributed equally to this work.}}
\author[a]{Shasha~Liu,\textsuperscript{\ref{co_author}}}

\author[a]{Ruike~Cao,}
\author[a]{Hengbin~Shao,}
\author[e]{Yaowei~Guo,}
\author[d]{Sirui~Wang,}
\author[a]{Fuyudi~Zhang,}
\author[a]{Haoyan~Zhang,}
\author[a]{Tailin~Zhu,}
\author[d]{Yixi~Jiang,}
\author[a,b]{Hao~Zhou,}

\author[a,b,c]{Xin~Xiang,\textsuperscript{\ref{co_correspondance}}}
\author[d]{Lei~Wang,\footnote[2]{\label{co_correspondance}Corresponding author(s). Email: xxiang@sjtu.edu.cn; wl@cdut.edu.cn.}}

\affiliation[a]{State Key Laboratory of Dark Matter Physics, Tsung-Dao Lee Institute \& School of Physics and Astronomy, Shanghai Jiao Tong University, Shanghai 200240, China}
\affiliation[b]{Key Laboratory for Particle Astrophysics and Cosmology (MOE) \& Shanghai Key Laboratory for Particle Physics and Cosmology, Shanghai Jiao Tong University, Shanghai 200240, China}
\affiliation[c]{Hainan Research Institute, Shanghai Jiao Tong University, Hainan 572024, China}

\affiliation[d]{College of Nuclear Technology and Automation Engineering, Chengdu University of Technology, Chengdu, China}

\affiliation[e]{Department of Physics, Imperial College of London, London United Kingdom}

\abstract{
We present the design and characterization of a Cockcroft--Walton (CW) high-voltage system developed for deep-sea neutrino telescopes. The system provides independently adjustable bias voltages to 31 three-inch PMTs inside a hybrid Digital Optical Module (hDOM). This paper describes the system design, control logic, test procedures, and the combined PMT–base performance, including baseline stability, gain uniformity, and timing accuracy. Performance was evaluated in the laboratory conditions that simulate the deep-sea environment. Baseline measurements indicate low and stable electronic noise, while gain calibration using single-photoelectron spectra shows that all PMTs can be tuned to a common nominal gain and remain stable over multi-day operation. Transit-time-spread measurements yield values below 1.8~ns (FWHM), consistent with manufacturer specifications. These results demonstrate that the CW-based system delivers the stability and timing precision required for deep-sea multi-PMT optical modules.
}

\keywords{PMT, Cockcroft-Walton HV System, Neutrino Telescope}

\arxivnumber{2512.14382} 

\begin{document}
\maketitle

\flushbottom

\section{Introduction}
\label{sec:intro}

Neutrino telescopes detect high-energy neutrinos from astrophysical sources such as active galactic nuclei and core-collapse supernovae, providing probes of cosmic accelerators, multi-messenger phenomena, and fundamental physics beyond the Standard Model~\cite{Gaisser2016, Halzen:2010yj, Adrian-Martinez2016}. Experiments like IceCube and KM3NeT primarily rely on the detection of Cherenkov photons produced by charged particles generated in neutrino interactions within transparent media such as seawater or ice. Photomultiplier tubes (PMTs) serve as the primary optical sensors, converting photon signals into electrical pulses, whose timing and charge encode the event topology and energy. Because the detectors are deployed at great depths or embedded in ice, the PMTs and their front-end electronics must operate reliably inside sealed, pressure-resistant glass vessels that preserve optical clarity and mechanical integrity. Stable and well-regulated high voltage is essential, as it directly determines the PMT gain and timing response, which in turn impacts photon hit resolution, and ultimately the quality of neutrino reconstruction. Therefore, the design of the PMT housing, its high-voltage, and readout systems plays a critical role in the overall sensitivity and long-term performance of the telescope.

The TRIDENT(TRopical DEep-sea Neutrino Telescope) neutrino telescope deploys an array of hybrid Digital Optical Modules (hDOMs) anchored at depths exceeding 3,000 m~\cite{TRIDENT:2022hql}. Each hDOM integrates two complementary photon-sensing technologies—31 three-inch PMTs and 24 silicon photomultipliers (SiPMs)—to achieve high photo-coverage and directional sensitivity~\cite{trident2023hdom, Shao2025_TRIDENT}. This hybrid approach follows the multi-PMT DOM concept pioneered in KM3NeT~\cite{KM3NeT_DOM_2018}. The sensors are housed within a 17-inch-diameter borosilicate glass vessel, manufactured by Nautilus, with a 14 mm wall thickness, providing over 95 \% optical transparency and sufficient mechanical strength to withstand pressures up to 670 bar \cite{VITROVEX17}. An aluminum support frame holds all sensors in fixed orientations, while the volume between the frame and the inner wall of the glass vessel is filled with high-transparency optical gel that ensures stable optical coupling. 

The compact integration of multiple sensors within the 17-inch vessel imposes stringent requirements on system stability and control. Because recovery or maintenance at kilometer-scale depths is costly and technically demanding, each hDOM must operate reliably for long periods without intervention. The high-voltage (HV) system is a critical component, as its stability directly determines the PMT gain and signal-to-noise ratio, which in turn affect the detection of faint Cherenkov photons and the accuracy of event reconstruction.To meet these requirements, a CW voltage multiplier is adopted for its compact and efficient design. Its ability to generate high voltages with a minimal number of components makes it particularly suitable for the confined geometry of deep-sea optical modules ~\cite{IceCube_DOM_2006}. 


This paper presents the design and performance evaluation of a CW based high-voltage system developed for deep-sea neutrino telescope applications~\cite{CockcroftWalton1932}. The system architecture is described with an emphasis on achieving low output ripple and long-term voltage stability. Performance is validated through circuit simulations and laboratory measurements. Sec.~\ref{sec:PMT_HV} outlines the CW circuit design, control logic, and test procedures. Sec.~\ref{sec:PMT_Performance} reports the PMT response, including baseline ripple and gain stability under simulated deep-sea conditions. Sec.~\ref{sec:Summary} summarizes the results and discusses their relevance to future large-scale neutrino detectors.


\section{The PMT HV System}
\label{sec:PMT_HV}

\subsection{Overall Design}

The HV system adopts a distributed architecture in which each PMT is equipped with an independent Cockcroft–Walton (CW) base. This one-CW-per-PMT scheme ensures uniform gain tuning, isolates channel-level variations, and avoids single-point failure modes. The system follows a three-tier hierarchy: the hDOM motherboard FPGA~\cite{Zhang:2025vvf} issues high-level commands; a dedicated control board manages command fan-out and channel selection; and the individual PMT bases execute voltage generation. Each base maintains a compact $47\times47$ mm footprint, allowing all 31 PMTs to be integrated within the restricted mechanical envelope of the deep-sea optical module.

\begin{figure}
    \centering

    \begin{subfigure}[t]{0.485\textwidth}
        \centering
        \adjustbox{valign=T}{%
            \includegraphics[width=\linewidth]{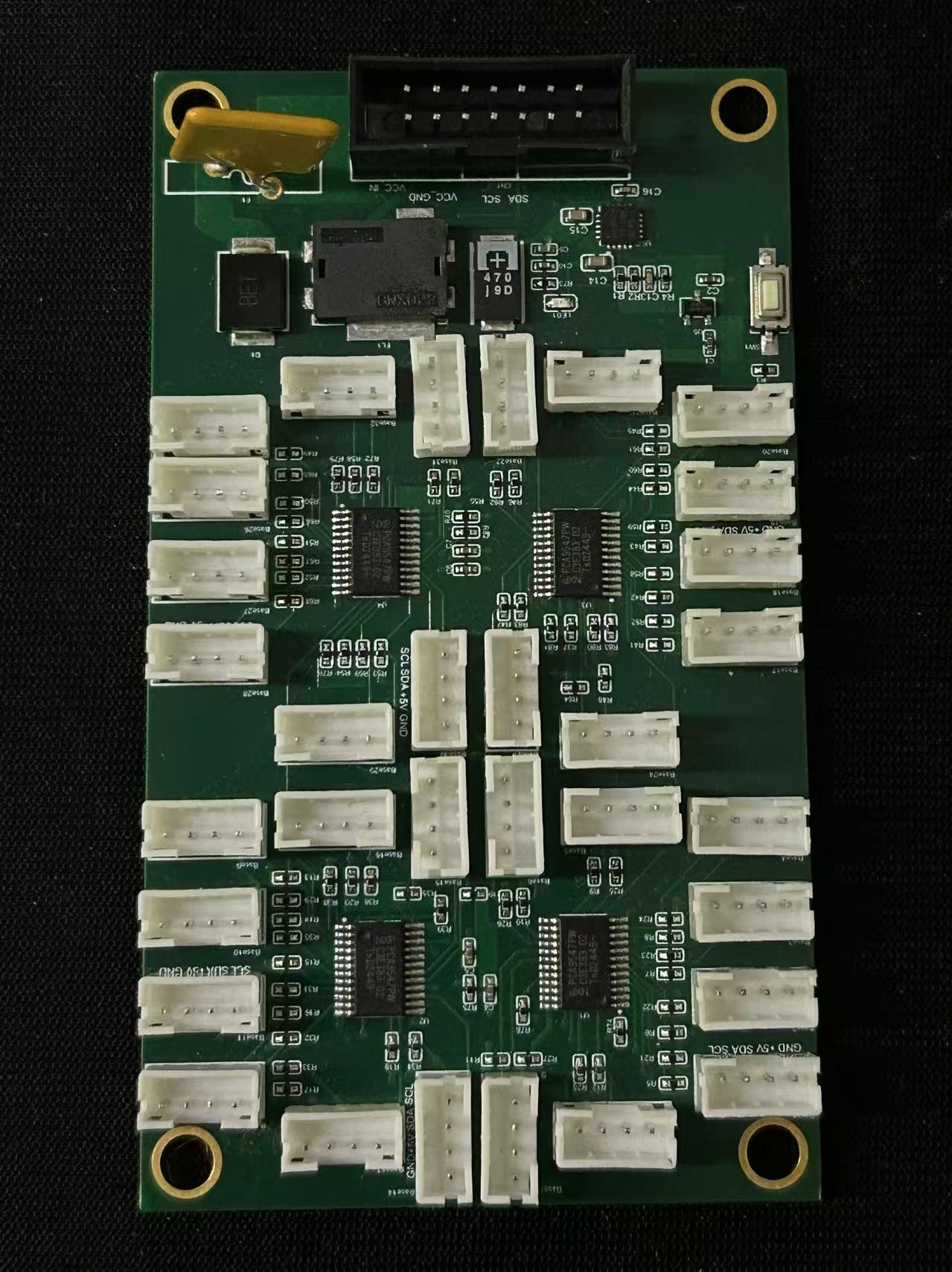}%
        }
    \end{subfigure} 
    \hfill 
    \begin{subfigure}[t]{0.5\textwidth}
        \centering
        \adjustbox{valign=T}{%
            \includegraphics[width=1.3\linewidth, angle=-90]{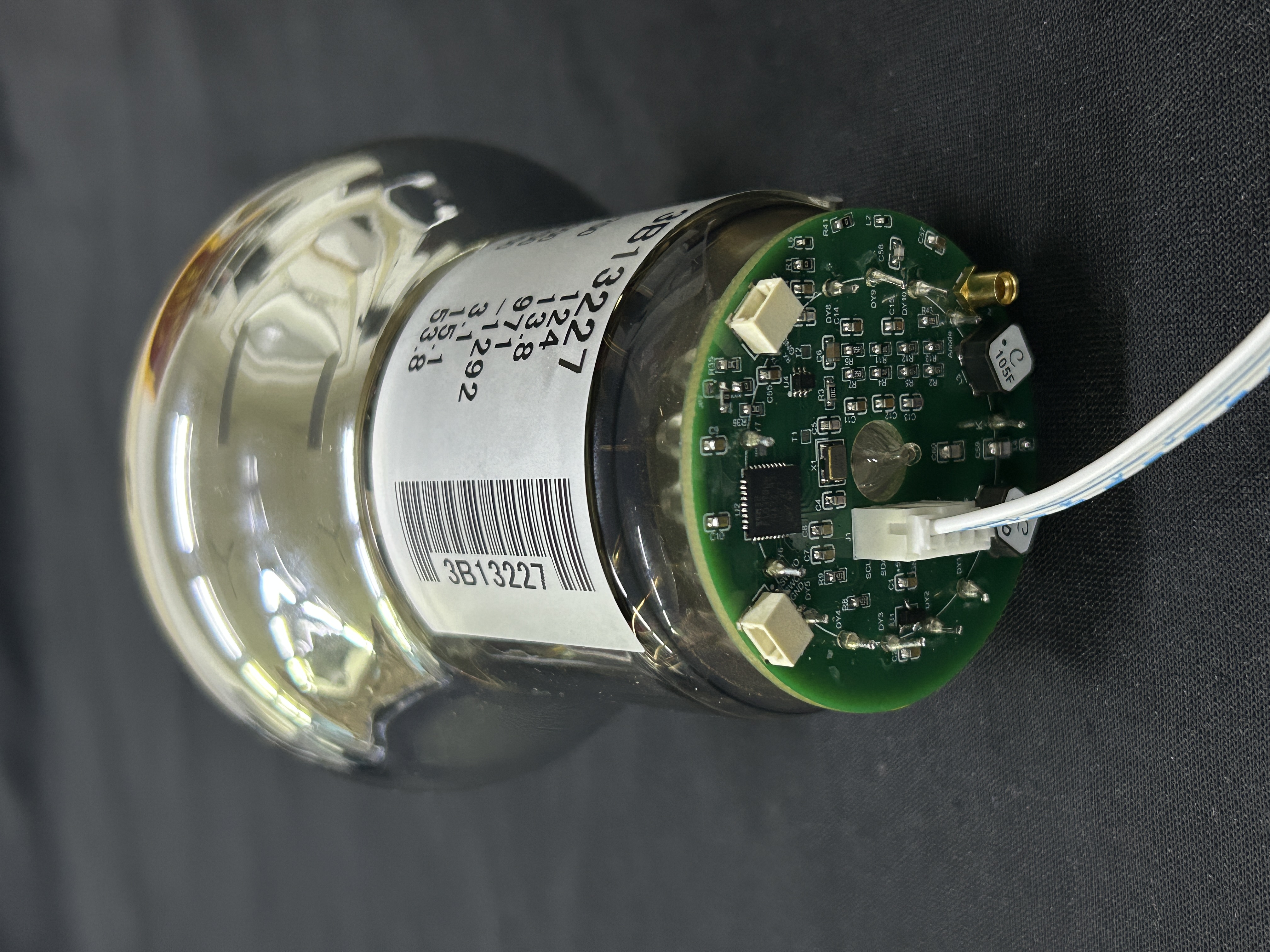}%
        }
    \end{subfigure} 
    
    \caption{Photo of the PMT high-voltage system, consisting of (left) the control board and (right) the PMT-integrated base board}
    \label{fig:block diagram of system}
\end{figure}

Fig.~\ref{fig:block diagram of system} shows the actual photos of the control and base boards. The FPGA motherboard digitizes PMT signals and communicates with the control board via a shared I\textsuperscript{2}C interface to adjust bias voltages. The control board, acting as an intermediate hub, distributes power and commands to the bases, simplifying the motherboard's layout. This work focuses specifically on the design and characterization of the control board and the per-PMT bases. The motherboard was also designed and developed by the TRIDENT collaboration, but it will not be discussed in detail here.


\subsection{Base and Control Board Implementation}

\begin{figure}
    \centering
        \includegraphics[width=\linewidth]{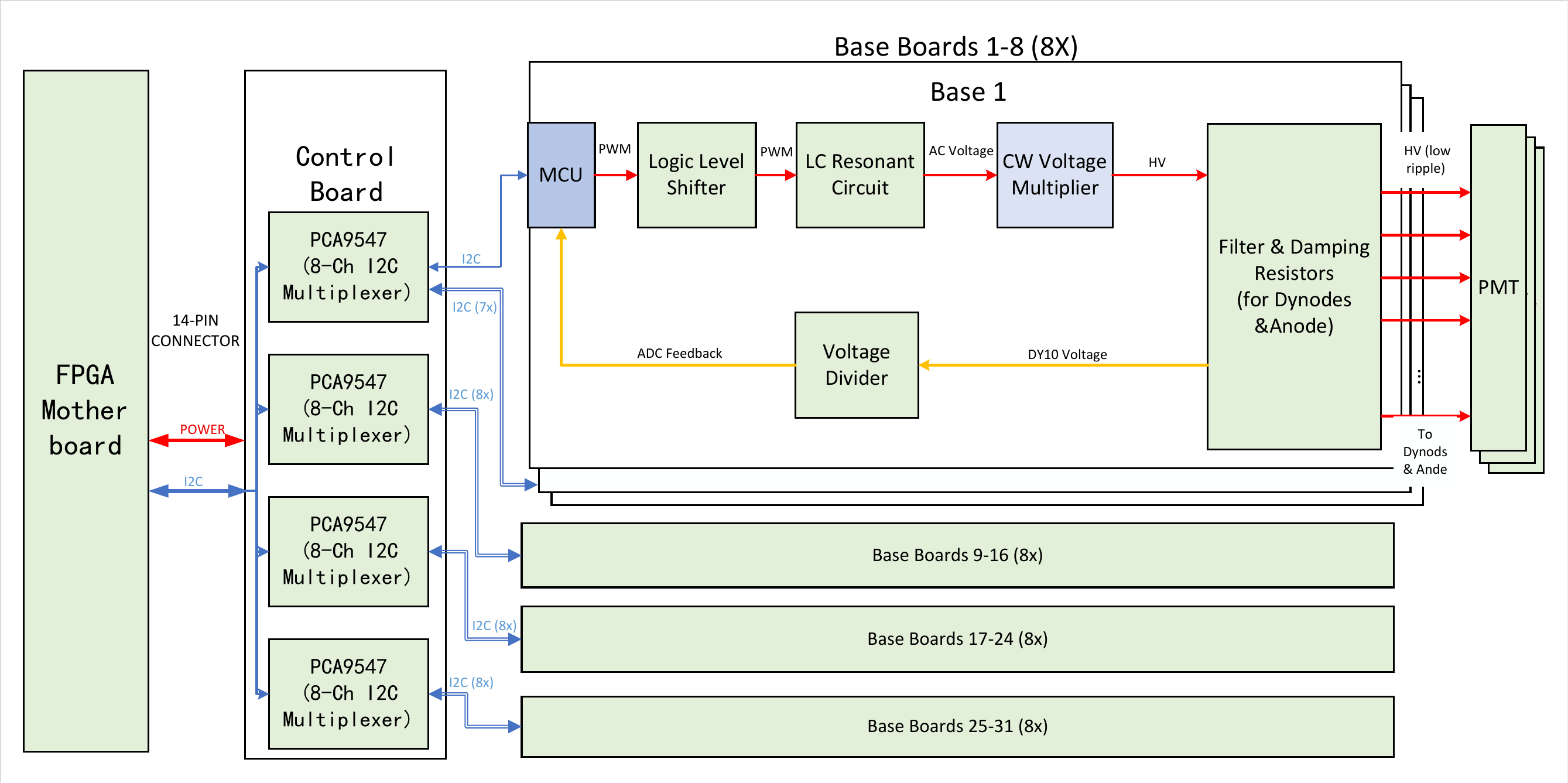}
        \caption{Schematic of the three-tier architecture for the PMT high-voltage system.}
        \label{fig:base circuit}
\end{figure}

The control board manages communication between the FPGA motherboard and the 31 bases through a multiplexed $I^2C$ architecture. It utilizes four 8-channel $I^2C$ multiplexers with reset capability (PCA9547) to interface with the FPGA via a 14-pin header. As illustrated in Fig.~\ref{fig:base circuit}, each multiplexer is routed to eight base connectors. To prevent bus contention arising from the PCA9547's default power-up state—where Channel 0 is active—the FPGA firmware initially disables all multiplexer channels. Communication is subsequently established by enabling only the specific channel corresponding to the target base, allowing for isolated voltage configuration and data readback. This architecture eliminates the requirement for 31 unique device addresses and has been experimentally verified to provide reliable, conflict-free monitoring across all channels.


Each base comprises three functional blocks: an MCU, an inductive resonant driver, and a Cockcroft-Walton (CW) voltage multiplier (Fig.~\ref{fig:base circuit}). The MCU generates a pulse-width–modulated (PWM) signal to drive the resonant stage, which employs inductors to produce an AC waveform. This waveform is subsequently amplified and rectified by the CW diode-capacitor network to provide an adjustable output from 0 V to $-1.5$ kV. Output regulation is achieved by tuning the PWM frequency and duty cycle. For closed-loop control, the high-voltage output is scaled by a precision resistor divider and digitized by the MCU’s internal ADC, providing real-time feedback for regulation.



The resonant driver and CW multiplier constitute the core high-voltage generation circuit. Depending on the resonant parameters, the PWM frequency is set to approximately 80 kHz, producing a 100–150 V${\text{pp}}$ AC signal that feeds the multiplier. The working principle of the CW multiplier can be understood by analyzing the two phases of the input AC waveform. During the positive half-cycle, the diodes are forward-biased to charge the upper-tier capacitors to the peak input voltage ($V_{\text{p}}$). Upon polarity reversal during the negative half-cycle, these capacitors act as series voltage sources that, combined with the input drive, charge the lower-tier capacitors. This results in a voltage of approximately $2 V_{\text{p}}$ at the first stage, with the multiplication effect propagating through subsequent cascaded stages to achieve the required high-voltage output. A key advantage of the CW topology is that the voltage stress on individual capacitors does not exceed $2 V_{\text{p}}$, permitting the use of capacitors with voltage ratings significantly lower than the total output. The polarity of the output is determined by the diode orientation; for example, a positive high voltage is obtained by reversing the diode directions relative to a negative-output configuration.

The multiplier is configured to provide a 3:1:1:1:1:1:1:1:1:1:1 dynode voltage ratio. This specific distribution is recommended for the Hamamatsu R14374 and NNVT N2031 PMTs to minimize transition-time spread and maintain timing uniformity. The high-voltage output is then routed through a filter network and series damping resistors. These resistors suppress overshoot and high-frequency ringing caused by the PMT’s output capacitance, improving pulse stability. A parallel RC filter smooths the rectified output to provide a stable DC bias for PMT operation. To ensure signal integrity, low/high-pass filters and ferrite beads are integrated within the driver stage to suppress switching noise~\cite{Timmer_2010, IceCube:2021_mDOM}. The bases operate inside a sealed 0.5–0.7 atm nitrogen environment within the glass sphere, where the seawater temperature remains 2–4 °C. Under these conditions, the design should produce stable high-voltage outputs with sub-millivolt baseline fluctuations.


\subsection{Base High Voltage Test}

\begin{figure}
    \centering
    \includegraphics[width=\textwidth]{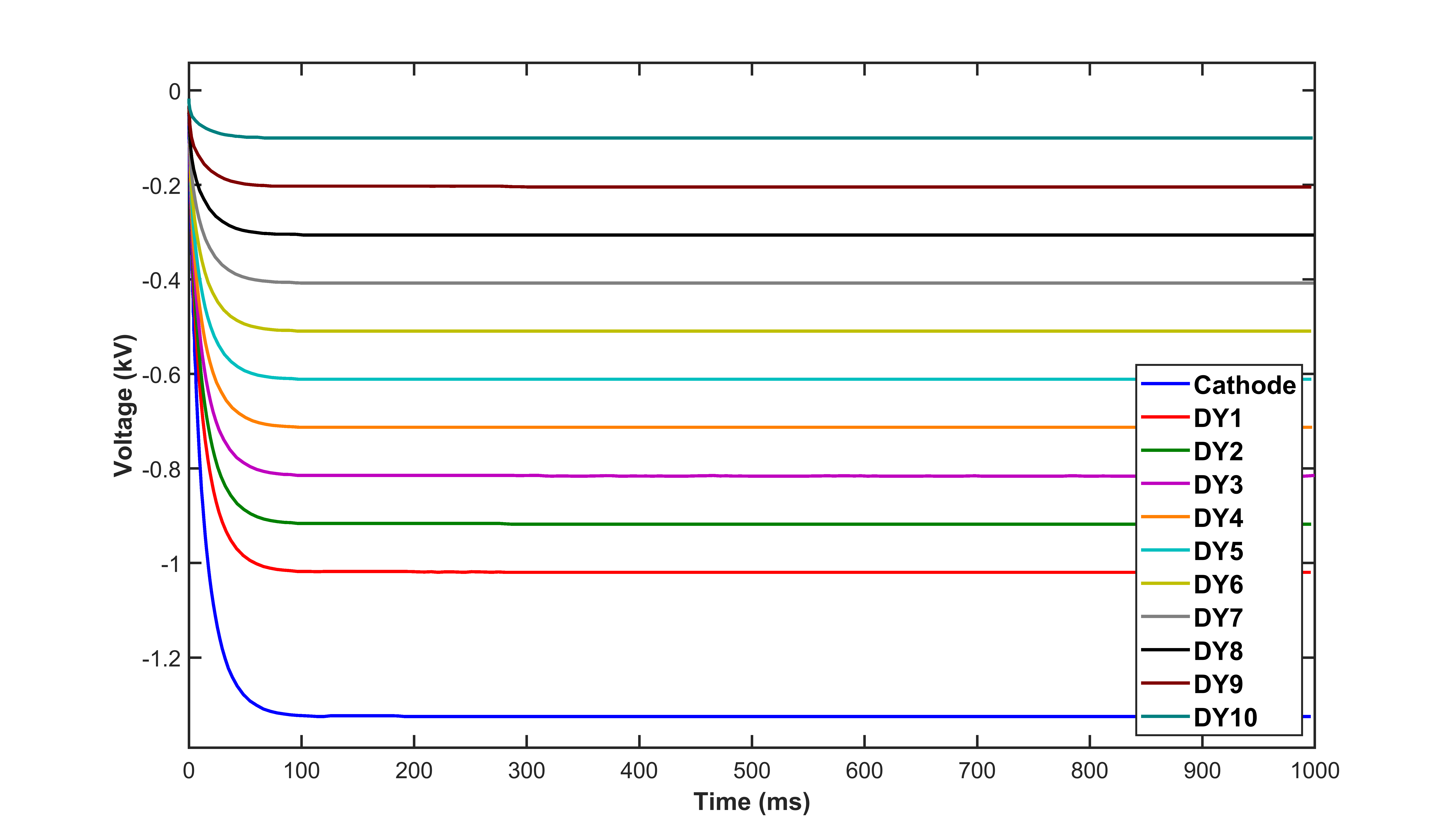} 
    \caption{Simulated high-voltage time evolution at different stages of the CW multiplier. All stages converge to their steady-state values with uniform voltage spacing along the multiplication chain.}
    \label{fig:Simulated high voltage}
\end{figure}

We simulated the performance of the PMT base to evaluate voltage buildup and steady-state behavior across the multiplier stages. The circuit was modeled in ADI LTspice using component values that match those of the prototype, including the resistor–capacitor network and output filter. Fig.~\ref{fig:Simulated high voltage} shows the simulated stage voltages as a function of time. All stages reach their steady-state values after roughly 100 ms, with uniform voltage spacing along the multiplication chain that is consistent with the design specifications.

\input{table/stage_voltages}

For laboratory characterization and firmware updates, a USB-to-serial interface was utilized to connect the base MCU directly to a host PC. This interface acts as a communication bridge, converting USB signals into the asynchronous serial data (UART) required by the MCU, allowing the base to be tested independently of the hDOM motherboard.

Using this setup, three bases were tested to verify dynode voltages and consistency with the designed ratio. Each unit was placed on an insulating substrate, and stage voltages from DY10 to the cathode were measured sequentially using a high-voltage probe (HVP-40, 1000:1 attenuation) after a 5-second stabilization period. The first base yielded -104~V at DY10, which was adopted as the reference setting for the remaining units. During measurements, the anode (P) was held at ground, and the cathode (K) reached approximately -1270~V. Table~\ref{tab:voltage_data} lists the measured stage voltages, which follow the expected CW scaling and agree with the LTspice simulation. Key performance results under a regulated 5~V input and a -1200~V output are summarized in Table~\ref{tab:hv_specs}.



\input{table/data_result}

\subsection{Control Logic}

The PMTs typically operate at -1250 V to achieve a gain of $10^{7}$. Since the gain follows a power-law dependence on the bias voltage, small variations in the bias can lead to significant gain deviations. The required high voltage varies between PMTs. Furthermore, the gain may drift over extended periods of operation, and specific channels may need to be selectively disabled under deep-sea conditions. These considerations necessitate fine, stable, and independently configurable high-voltage control for all PMTs within an hDOM.

\begin{figure}
    \centering
    \includegraphics[width=\textwidth]{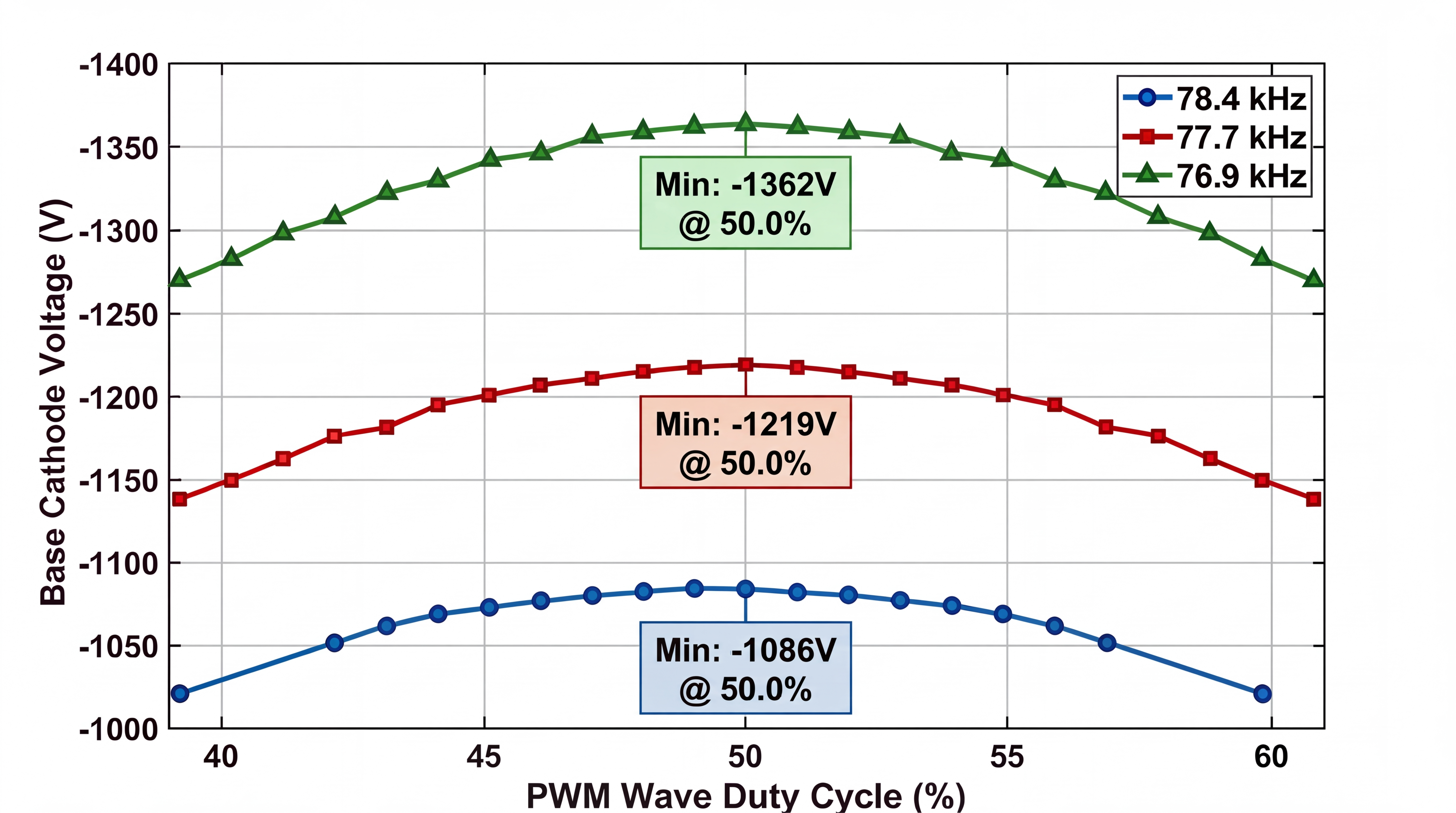}
    \caption{Cathode voltage as a function of the duty cycle at different driving frequencies}
    \label{fig:control_logic}
\end{figure}

We conducted a systematic sweep of the PWM parameters using the laboratory bench setup to map the circuit's transfer function. With the base connected to a host PC, the MCU was commanded to sweep through a range of frequencies and duty cycles. The resulting high-voltage output was sampled by the on-board ADC and transmitted back to the PC. As shown in Fig.~\ref{fig:control_logic}, the relationship between PWM parameters and the cathode voltage exhibits resonant behavior. At a fixed frequency, the output peaks near a 50\% duty cycle and follows a near-parabolic profile. Adjusting the PWM frequency provides coarse control—shifting the resonant peak to allow for approximately 100~V steps while fine-tuning the duty cycle enables precision regulation.




After system integration, the FPGA executes a two-step autonomous control sequence. The FPGA first scans the PWM drive frequency around an initial reference to reach the approximate target range, and then iteratively fine-tunes the duty cycle based on feedback from the on-board ADC until the desired high-voltage value is achieved. This closed-loop procedure enables each PMT base to autonomously converge to its operating voltage with sub-percent precision, ensuring uniform gain calibration and stable performance across all channels.

\section{PMT Performance}
\label{sec:PMT_Performance}

\subsection{Experimental Setup}
\label{sec:Experimental_Setup}
 
\begin{figure}
    \centering
    \includegraphics[width=\textwidth]{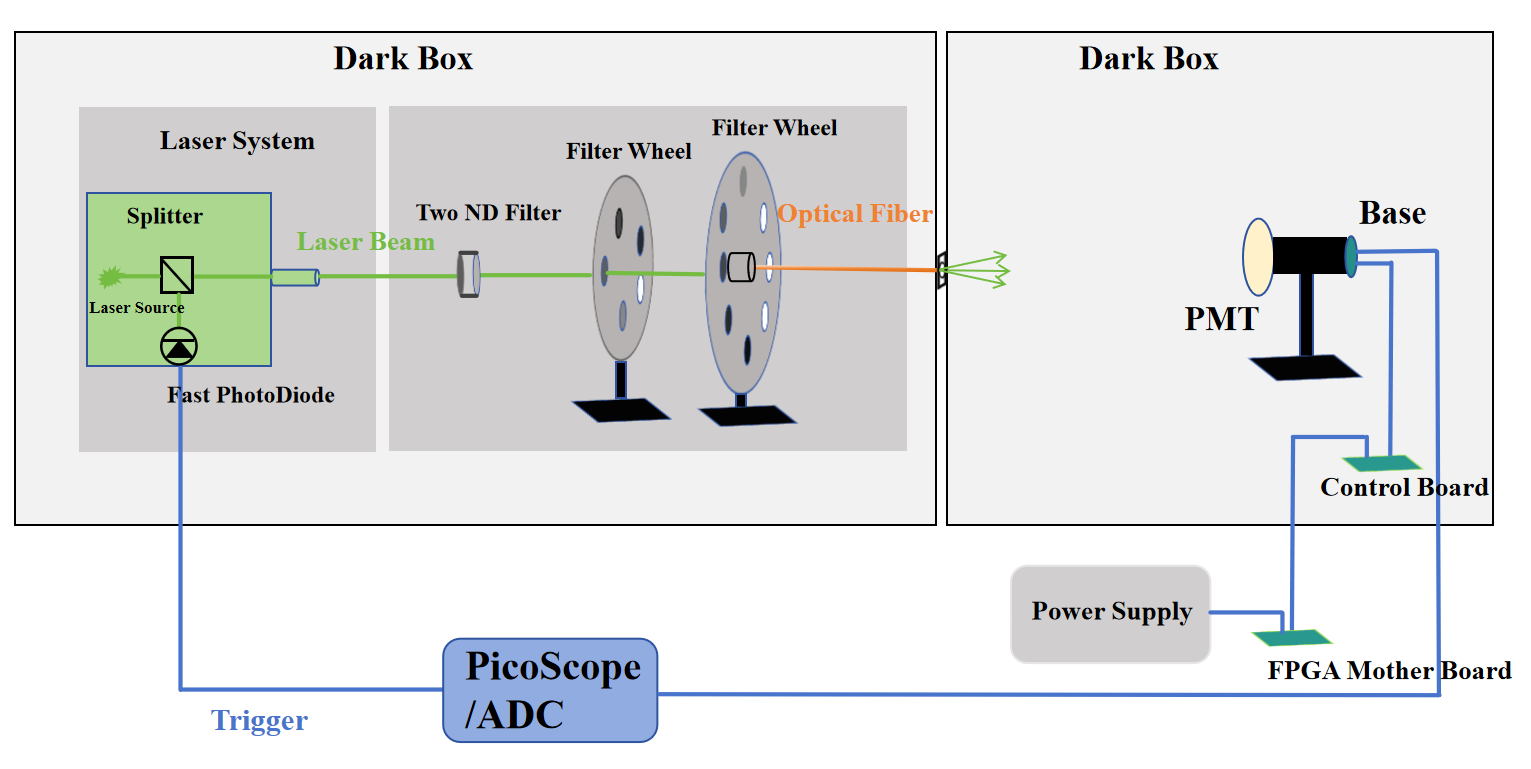}
    \caption{The PMT bench test setup designed for SPE gain,gain curve and time resolution measurements with main components marked.}
    \label{fig:PMT_setup}
\end{figure}

The performance of the PMTs with the bases is directly linked to the physical performance of the hDOM detector. We evaluate the stability and timing characteristics of the integrated units through two distinct experimental configurations:

\begin{itemize}
\item \textbf{Laser Test Bench (Individual Characterization):} Used to measure parameters such as the gain-voltage curve and Transit Time Spread (TTS). As illustrated in Fig.~\ref{fig:PMT_setup}, a picosecond laser ($532$~nm, $\sim 10$~ps pulse width) with an adjustable repetition rate ($500$~Hz to $300$~kHz) served as the light source. The laser output was split into a reference trigger branch, coupled to a fast photodiode, and an illumination branch. To reach the single-photoelectron (SPE) regime, the illumination passed through motorized neutral-density (ND) filter wheels (Tab.~\ref{tab:ND}) and a diffuser fiber tip. Data were captured using a PicoScope 6426E ($5$~GS/s sampling rate, $1$~GHz bandwidth) triggered externally by the laser synchronization signal. This ensures that incident photons arrive at a fixed, predictable time delay relative to the trigger, facilitating high-precision timing analysis.
\item \textbf{Integrated hDOM Setup (System-Level Validation):} Used to assess long-term gain and baseline stability for the full array of 31 PMTs (12 NNVT, 19 Hamamatsu) integrated with the complete HV system. To simulate deep-sea conditions, the hDOM was placed in a climatic chamber at $2^{\circ}\mathrm{C}$ and backfilled with dry nitrogen to $\sim 0.5$~atm. Data acquisition in this setup supports three distinct trigger modes: (1) PMT self-trigger, which initiates a readout when a single-channel pulse height exceeds a predefined threshold; (2) periodic random trigger, used primarily for unbiased sampling of the baseline and dark count rate (DCR); and (3) local coincidence trigger (CL2), which initiates a readout when two or more PMTs exceed their thresholds within a 20~ns coincidence window. These modes allowed the FPGA motherboard to continuously monitor the integrated HV system’s stability and noise floor over a 100-hour period.
\end{itemize}

Both configurations were housed in dark rooms to maintain a stable optical environment. The PMTs were granted an adequate stabilization period to reach thermal and electronic equilibrium, ensuring the DCR had settled near the manufacturer's reference value ($\sim$1~kHz) prior to data acquisition.

\input{table/ND}

\subsection{PMT gain}
\label{sec:PMT gain}

The PMT gain was calibrated using the laser test bench to characterize the response of individual units. The procedure involves measuring single-photoelectron (SPE) charge spectra under low-light conditions, where the mean photoelectron occupancy satisfies $\lambda \ll 1$. The charge of each PMT pulse is obtained by integrating the waveform over a predefined time window around the pulse. The resulting charge distribution, $f(q)$, can be modeled as a Poisson-weighted sum of $n$-photoelectron contributions \cite{DEAP:2017fgw}:
    \begin{equation}
    \label{eq:full-model}
    \begin{aligned}
    f(q)
    =~& B\Bigl(
    \operatorname{Poisson}(0;\lambda)\,\mathrm{Ped}(q)
    + \sum_{n=1}^{\infty} \operatorname{Poisson}(n;\lambda)\,
    \bigl(\mathrm{Ped} \otimes
    \underbrace{\mathrm{SPE}\otimes\cdots\otimes \mathrm{SPE}}_{n~\text{times}}
    \bigr)(q)\Bigr).
    \end{aligned}
    \end{equation}   

where $\mathrm{Ped}(q)$ denotes the Gaussian electronic pedestal with two parameters $\mu_{\rm ped}$ and $\sigma_{\rm ped}$. $B$ is an overall
normalization factor, and $\lambda\ll 1$ is the mean photoelectron occupancy.

The SPE charge response is modeled as a
mixture of two Gamma components and an exponential low-charge tail~\cite{DEAP:2017fgw}:
\begin{equation}
\mathrm{SPE}(q) =
\eta_{1}\,\Gamma(q;\mu,b)
+ \eta_{2}\,\Gamma(q;\mu f_{\mu},\, b f_{b})
+ 
\begin{cases}
\eta_{3}\,\ell\,e^{-\ell q}, & q < \mu, \\[2pt]
0, & q \ge \mu ,
\end{cases}
\label{eq:spe}
\end{equation}
where $\eta_{1,2,3}\ge 0$ are the component weights satisfying
$\eta_{1}+\eta_{2}+\eta_{3}=1$.
The first Gamma term describes the full dynode multiplication process and is characterized by the mean SPE charge $\mu$ and a shape parameter $b$ that controls the width of the full-multiplication peak~\cite{PRESCOTT1966173}. The second Gamma term captures reduced-charge events from incomplete multiplication at early dynodes; its mean and shape are scaled relative to the primary component by $f_{\mu}$ and $f_{b}$, respectively. The exponential component, with decay constant $\ell$, models the continuous low-charge population arising from elastic backscattering and other partial-amplification processes. This Gamma-family parameterization has been shown to reproduce the full SPE lineshape in many cases~\cite{DEAP:2017fgw, Akashi-Ronquest:2019mlk, Kalousis:2020PMTCalib}.

\begin{figure}[!t]
  \centering
  \begin{subfigure}[t]{0.5\linewidth}
    \centering
    \includegraphics[width=\linewidth]{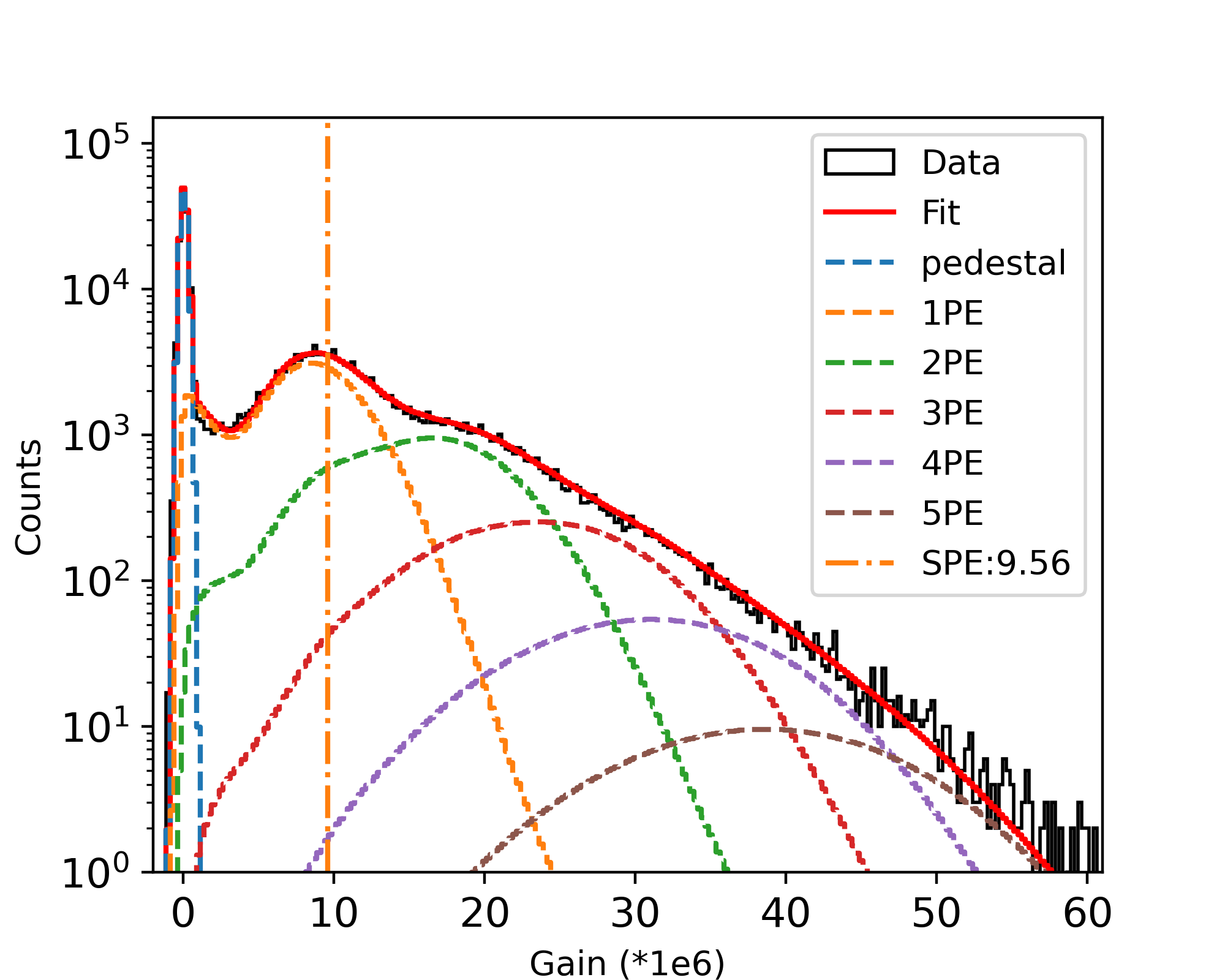}
    \label{fig:spe_fit}
  \end{subfigure}\hfill
  \begin{subfigure}[t]{0.5\linewidth}
    \centering
    \includegraphics[width=\linewidth]{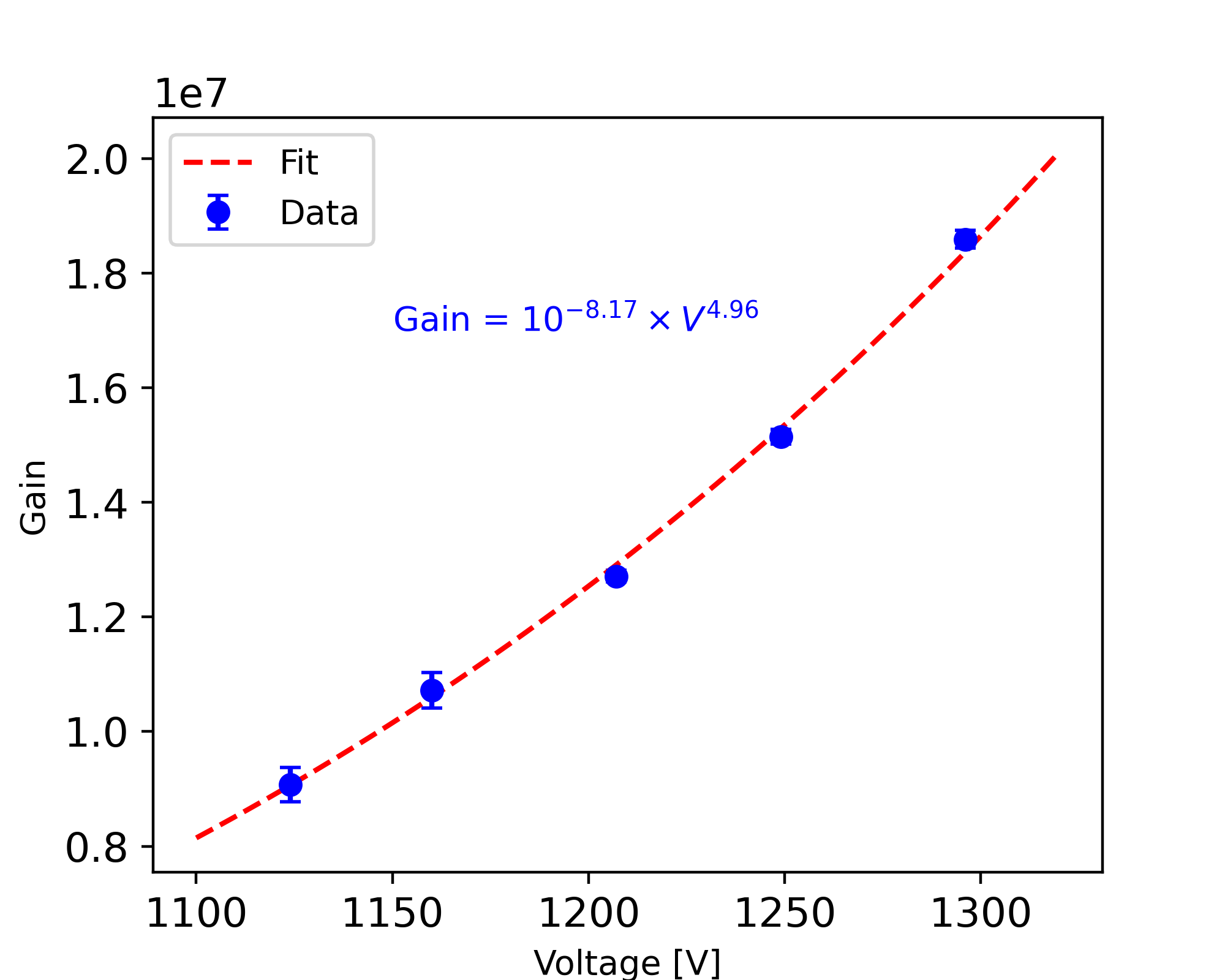}
    \label{fig:gain_curve}
  \end{subfigure}
  \caption{Left: The charge spectrum and the fit function. The measurement was performed with the experimental setup presented in Fig.\ref{fig:PMT_setup}. Right: SPE gain curve as a function of bias voltage. The points correspond to the mean charge of the 1-PE peak obtained from the fits in the left panel.}
  \label{fig:spe_gain}
\end{figure}

The overall fit function (Eq.~\ref{eq:full-model}) is governed by a total of 10 free
parameters: two pedestal parameters ($\mu_{\rm ped}$, $\sigma_{\rm ped}$),
seven SPE-response parameters ($\mu$, $b$, $f_{\mu}$, $f_{b}$, $\eta_{2}$,
$\eta_{3}$, $\ell$) with $\eta_{1}=1-\eta_{2}-\eta_{3}$ fixed by normalization,
and one Poisson occupancy parameter ($\lambda$). A binned $\chi^{2}$ minimization was used to extract the ten free parameters governing the pedestal, SPE response, and Poisson occupancy. An example fitted charge spectrum is shown in Fig.~\ref{fig:spe_gain} left. The gain uncertainty is obtained by propagating the parameter covariance from the SPE-spectrum fit to the mean SPE charge. This provides the statistical error on the extracted gain.


The dependence of gain on the bias voltage is well described by a power-law model,
\begin{equation}
G(V) = \alpha V^{\beta}
\label{gain_curve}
\end{equation}
where $\beta$ characterizes the voltage sensitivity of the amplification. To determine the parameters $\alpha$ and $\beta$, the gain was measured at several high-voltage settings between $-1100$~V and $-1300$~V. The resulting gain–voltage relation is shown in Fig.~\ref{fig:spe_gain} right. Because the gain–voltage response varies from PMT to PMT, measuring the gain curve for each unit allows the bias voltage to be tuned later to restore the target gain as the tube ages or its response drifts.

\begin{figure}[H]
    \centering
    \includegraphics[width=\linewidth]{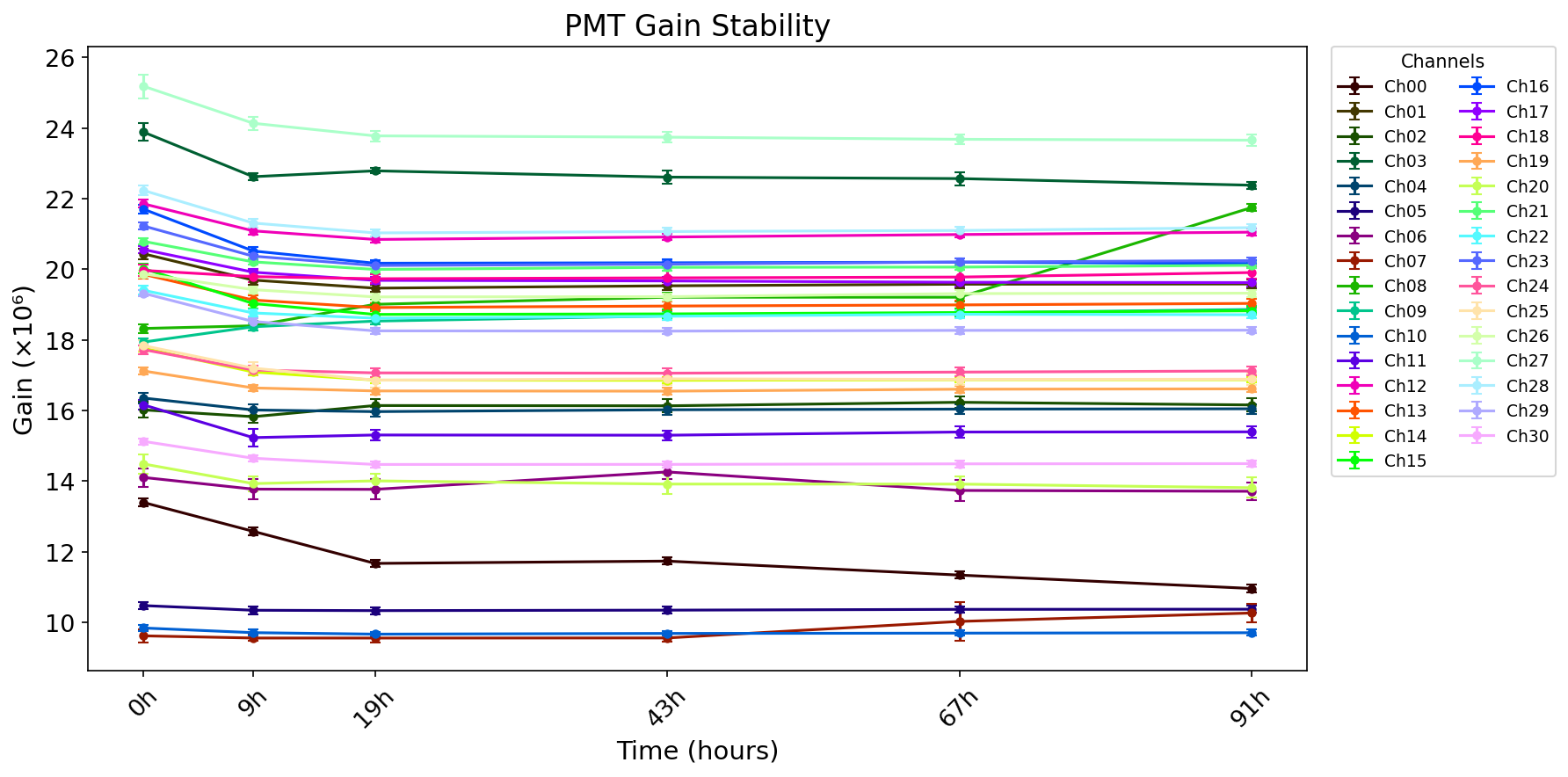}
    \caption{Gain stability of the 31 PMTs in the hDOM as a function of time. Each data point represents the mean gain over a 5-hour run, where the error bars indicate the covariance of the spe gain fit. The first run is defined as 0 h, and subsequent runs correspond to 9 h, 19 h, 43 h, 67 h, and 91 h, respectively. The gains stabilize after approximately 20 hours, with most PMTs fluctuating within $\pm2 - 3\%$ of their initial values. }

    \label{fig:gain_stability}
\end{figure}

\subsection{Gain and Baseline Stability}
\label{sec:gain stability}

The stability measurements were conducted using the integrated hDOM setup. Following the assembly of the 31 PMTs into a single hDOM, the hDOM was continuously powered and operated for around 100 hours at a fixed high voltage (corresponding to a gain of $1\times10^{7}$) to assess their long-term gain and baseline stability. Throughout this period, six data runs, each covering around 5~hours, were recorded using the motherboard designed for TRIDENT Phase-I~\cite{Zhang:2025vvf, yyong2025}. The motherboard is controlled by a host PC via a high-speed optical fiber link. Both power and communication are delivered through an electro-optical cable connected to the hDOM via a specialized titanium feedthrough penetrator in the glass vessel.

After determining the gain–voltage parameters (Sec.~\ref{sec:PMT gain}), the bias of each PMT was set to achieve a uniform nominal gain. The long-term gain stability of the 31 integrated PMTs was then evaluated using dark-count data, with each recorded waveform spanning 480~ns (60 ADC samples). The initial run defines $t = 0$~h, with subsequent measurements taken at 9, 19, 43, 67, and 91~h. Each data point in Fig.~\ref{fig:gain_stability} represents the mean gain extracted from a 5-hour dataset, where the error bars denote the uncertainty derived from the SPE-fit covariance. The gains are observed to stabilize after approximately 20~hours of continuous operation. For most channels, the fluctuations remain within $\pm 2$--$3\%$ of the initial values with no observable systematic drift. Even the channels with the highest gain settings maintain stability within 5\% throughout the 100-hour test period.

\begin{figure}[!t]
    \centering
    \includegraphics[width=\textwidth]{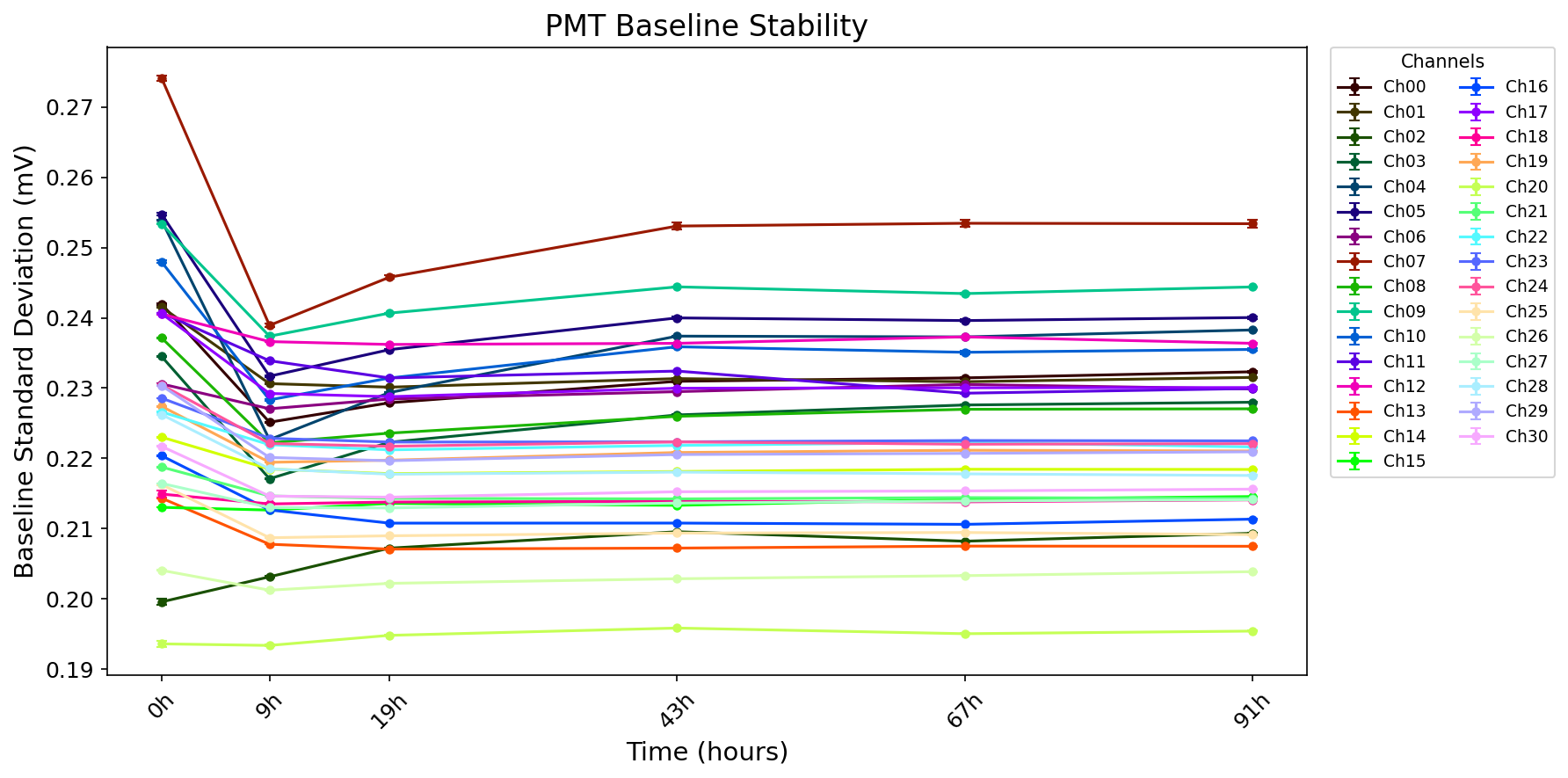}
    \caption{
    Baseline fluctuation of 31 PMT channels as a function of operation time. The baseline values were derived from the dataset described in Fig.~\ref{fig:gain_stability}, each data point represents the mean baseline fluctuation over a 5-hour run, where the error bars indicating the standard deviation within that period. The results show that the baseline stabilizes after about 40 hours of continuous operation, with channel-to-channel variations constrained within $2\%$, demonstrating negligible drift in front-end electronic noise.}
    \label{fig:baseline_stability}
\end{figure}

The baseline stability analysis was conducted using the same dataset as the gain stability study. The standard deviation of the first 20 samples (160~ns) of each waveform was used to quantify noise fluctuations, with the error bars representing the standard deviation of these fluctuations within each run. This interval was selected from the pre-trigger buffer, a region preceding the arrival of the PMT pulse, ensuring that the estimate reflects the intrinsic electronic noise. As shown in Fig.~\ref{fig:baseline_stability}, the baseline reaches a steady state after roughly 40~h of continuous operation, with channel-to-channel variations remaining within 2\%. This indicates negligible drift in the front-end electronic noise once thermal equilibrium is established.

\begin{figure}[!t]
    \centering
    \includegraphics[width=\textwidth]{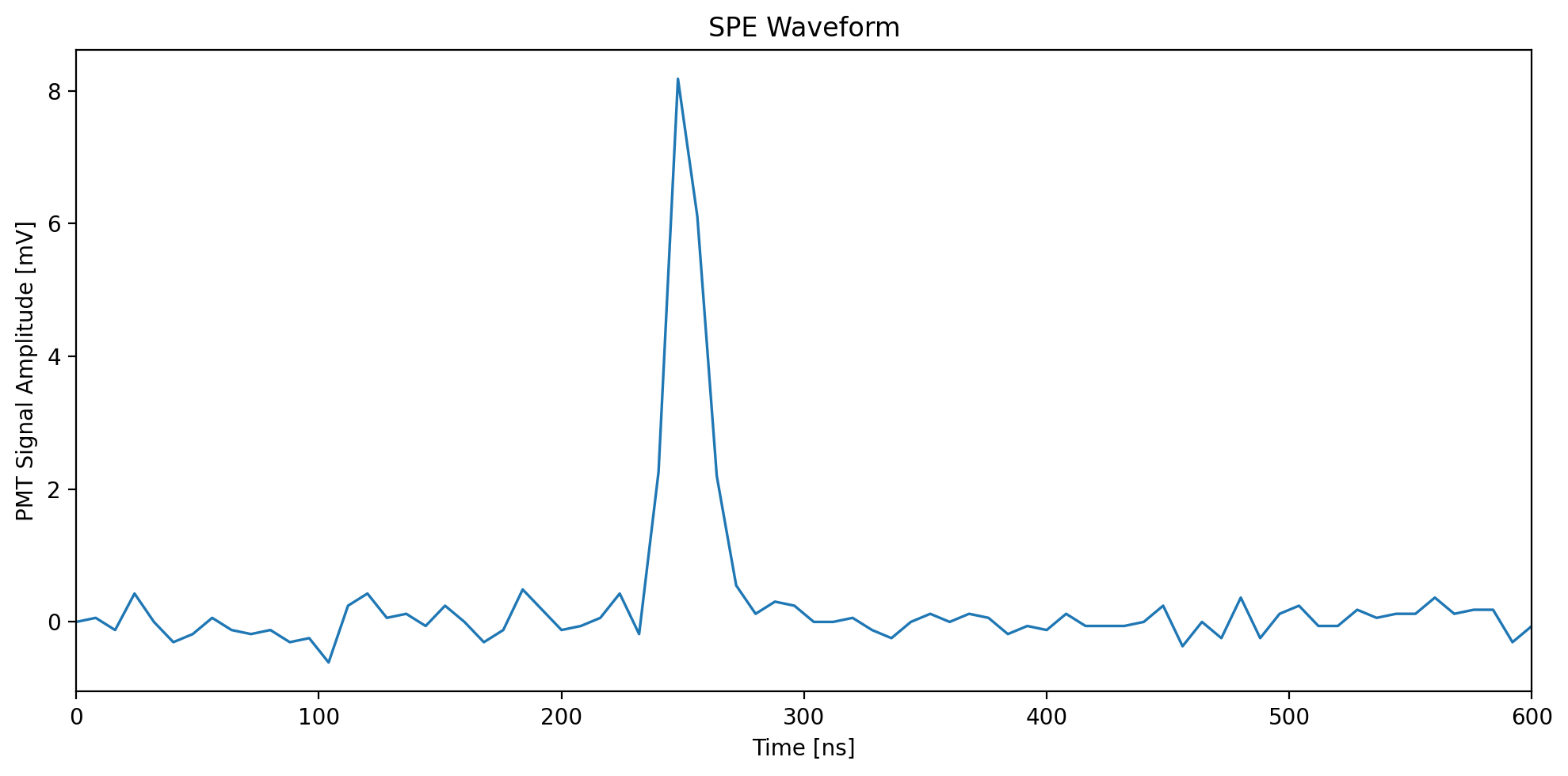}
    \caption{Representative SPE waveform at $10^{7}$ gain recorded by the integrated hDOM setup.}
    \label{fig:spe_waveform}
\end{figure}

The Signal-to-Noise Ratio (SNR) was evaluated by comparing the SPE amplitude to the baseline noise. This ratio is critical for high detection efficiency and low false-trigger rates. Fig.~\ref{fig:spe_waveform} shows a representative SPE waveform at a gain of $10^{7}$. The pulse peak is $\sim 8.2$~mV against a baseline RMS of $\sim 0.2$~mV, yielding an SNR of $\sim 40$. This high signal fidelity ensures a stable trigger threshold, meeting TRIDENT requirements for long-term deep-sea operation.

\subsection{PMT Time Resolution}

Precise photon timing is essential for neutrino telescopes, as the reconstruction of tracks and cascade directions and vertices relies on the arrival times of Cherenkov photons at optical modules. Simulation has shown that sub-degree angular resolution for muon tracks can be achieved only if the per-photon timing precision is on the order of 2 ns~\cite{Aguilar2011_ANTARES}. This requirement arises primarily from the intrinsic transit-time spread (TTS) of the PMT, the chromatic dispersion in seawater, and the undersea acoustic positioning accuracy. The PMT TTS, defined as the full width at half maximum (FWHM) of the distribution of single photoelectron (PE) transit times, is a key detector parameter to be evaluated.

\begin{figure}[!t]
    \centering
    \includegraphics[width=\textwidth]{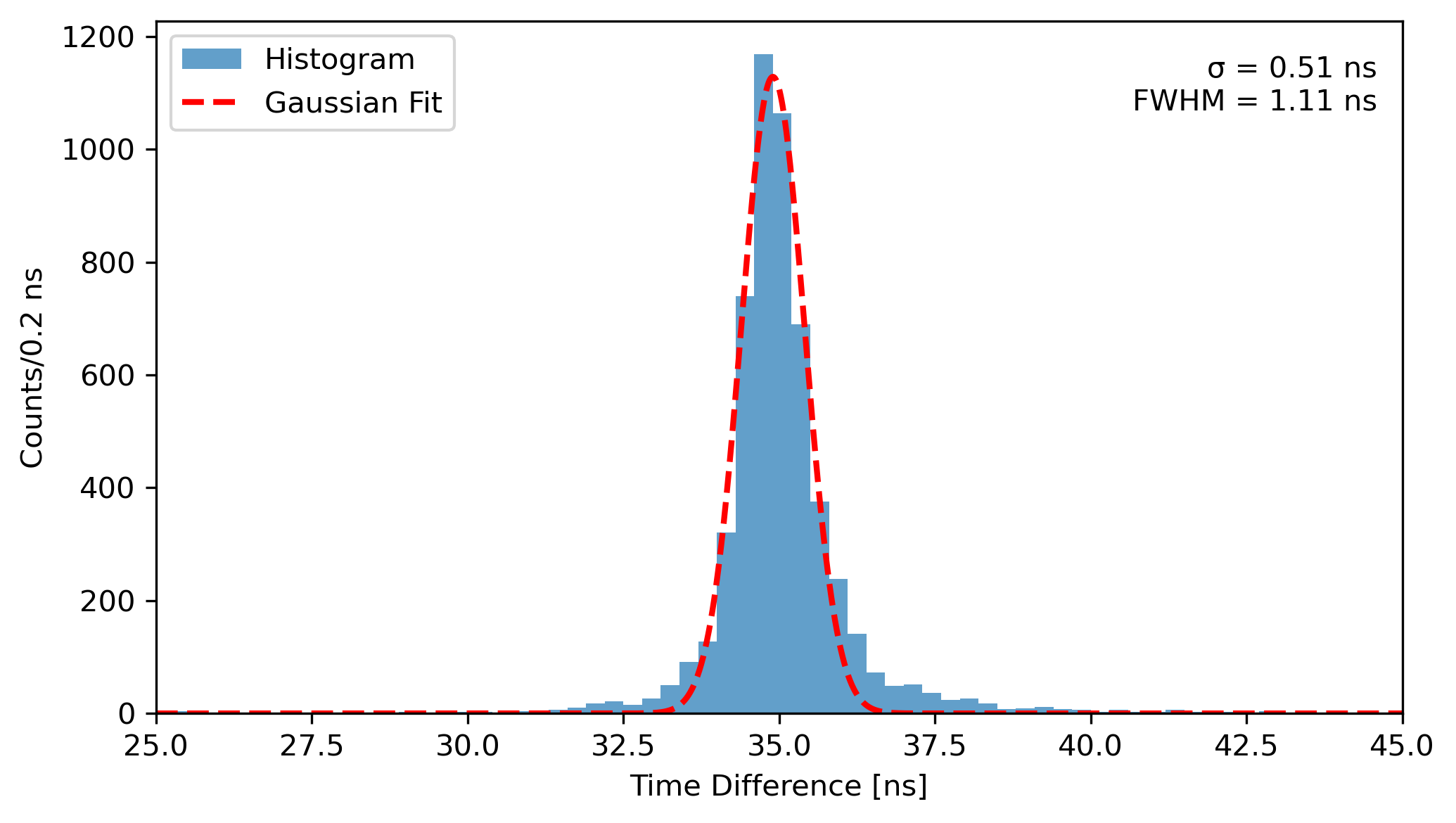}
    \caption{Transit Time Spread (TTS) measurement at a PMT gain of $10^7$. The histogram shows the distribution of the time difference between the laser trigger and the PMT response, and the red dashed line shows the Gaussian fit applied to the prompt peak. The FWHM extracted from the distribution is \SI{1.11}{\nano\second}.
}
    \label{fig:tts}
\end{figure}

The TTS of each PMT was measured using the laser test bench (Sec.~\ref{sec:Experimental_Setup}) with the intensity adjusted to maintain a low SPE occupancy ($\lambda \approx 0.05$). PMT waveforms were digitized by a PicoScope at 5~GSps, providing 200~ps temporal binning. The timing jitter of the photodiode reference channel is less than 20~ps, as calibrated by the manufacturer. This configuration allows sub-ns accuracy in the TTS measurement. 

The photon arrival time is defined as the point where the PMT waveform exceeds one-third of the averaged SPE peak amplitude. For each event, the transit time (TT) is calculated as the time difference between the PMT signal and the laser trigger. A fixed offset, arising from unequal optical and electrical propagation paths, was measured using a high-speed Silicon Photomultiplier (SiPM, SPTR~$\sim$150~ps ~\cite{trident2024sipm}) and subtracted from the results. 

A representative TTS measurement is shown in Fig.~\ref{fig:tts}. The TTS is typically extracted either from the width of the Gaussian core or directly from the full distribution. As expected for small-mass PMTs, the distribution shows a mild asymmetry with a late-time tail arising from delayed electron trajectories, including elastic back-scattering at the first dynode.

The Hamamatsu PMTs are specified to have a TTS below 1.8 ns (FWHM). This value was verified by the manufacturer using a conventional resistive-divider base during quality assurance tests. The measurements obtained with the CW-base configuration are consistent with these specifications, confirming that the CW design preserves the intrinsic timing performance required for deep-sea neutrino telescope applications.



\section{Summary and Discussion}

\label{sec:Summary}

This work presents the design, implementation, and validation of a Cockcroft–Walton–based high-voltage system developed for deep-sea neutrino telescopes. The system integrates an FPGA-controlled PWM driver, an LC resonant stage, and a multi-stage CW multiplier to provide independently adjustable, low-noise bias voltages for 31 PMTs within an hDOM. Laboratory tests demonstrate stable voltage regulation, accurate gain tuning, and reliable time performance over extended operation.

Baseline measurements show that all 31 PMTs maintain fluctuations at the $0.3\%$–$1.4\%$ level over five days, confirming the low-noise behavior of the front-end electronics. The CW-base configuration achieves a transit-time spread below 1.8~ns (FWHM), consistent with manufacturer specifications and sufficient for photon-timing requirements in deep-sea reconstruction. Gain measurements indicate that each PMT can be tuned to a common nominal gain, and the gain remains stable to within a few percent over the multi-day test period.

These results confirm that the system achieves the voltage stability, timing performance, and gain uniformity required for a deep-sea optical module, and is suitable for scaling to larger multi-PMT detectors in long-term deployments.

\acknowledgments

This work is supported by the Ministry of Science and Technology of China under National key research and development plan (Grant no. 2023YFC3107402). X. Xiang would like to thank the Double First-Class startup funds provided by Shanghai Jiao Tong University.







\bibliographystyle{JHEP}
\bibliography{ref}





\end{document}

%% file: table/stage_voltages.tex
\setlength{\arrayrulewidth}{0.5pt}
\renewcommand{\arraystretch}{1.2}

\begin{table}[H]
\centering
\small
\setlength{\tabcolsep}{4pt} 
\caption{Electrode Voltages of the PMT Base Board (V)}
\label{tab:voltage_data}
\begin{tabular}{>{\bfseries}l 
                >{\centering\arraybackslash}m{0.7cm}
                >{\centering\arraybackslash}m{0.7cm}
                >{\centering\arraybackslash}m{0.7cm}
                >{\centering\arraybackslash}m{0.7cm}
                >{\centering\arraybackslash}m{0.7cm}
                >{\centering\arraybackslash}m{0.7cm}
                >{\centering\arraybackslash}m{0.7cm}
                >{\centering\arraybackslash}m{0.7cm}
                >{\centering\arraybackslash}m{0.7cm}
                >{\centering\arraybackslash}m{0.7cm}
                >{\centering\arraybackslash}m{0.9cm}
                >{\centering\arraybackslash}m{0.9cm}}
\toprule
\textbf{Electrodes} & \textbf{P} & \textbf{Dy10} & \textbf{Dy9} & \textbf{Dy8} & \textbf{Dy7} & \textbf{Dy6} & \textbf{Dy5} & \textbf{Dy4} & \textbf{Dy3} & \textbf{Dy2} & \textbf{Dy1} & \textbf{K} \\
\midrule
V$_\text{Base1}$  & 0 & -104 & -208 & -311 & -412 & -513 & -615 & -706 & -805 & -900 & -998 & -1274 \\
V$_\text{Base2}$  & 0 & -104 & -208 & -312 & -415 & -515 & -613 & -711 & -806 & -902 & -997 & -1270 \\
V$_\text{Base3}$  & 0 & -104 & -209 & -312 & -414 & -515 & -613 & -711 & -806 & -900 & -995 & -1266 \\
V$_\text{sim}$  & 0 & -104 & -208 & -312 & -416 & -520 & -624 & -724 & -832 & -936 & -1040 & -1352 \\
\bottomrule
\end{tabular}
\end{table}

%% file: table/data_result.tex
\begin{table}[H]
\centering
\caption{Specifications of the High-Voltage Base board}
\label{tab:hv_specs}
\begin{tabular}{ll}
\toprule
\textbf{Parameter} & \textbf{Value} \\
\midrule
Input Voltage & $5 \text{ V}$ \\
Output Voltage Range & $-1500 \text{ V}$ to $0 \text{ V}$ \\
Ripple (Anode to GND) & $\sim 150 \text{ mV}_{\text{pp}}$ \\
Power Consumption (Switch-on) & $< 60 \text{ mW}$ (at $-1200 \text{ V}$) \\
Settling Time & $\sim 3 \text{ s}$ \\
Output Stability & $0.1\%$ (under stable input) \\
\bottomrule
\end{tabular}
\end{table}

%% file: table/ND.tex
\begin{table}[H]
  \centering

  \renewcommand{\arraystretch}{0.85}

  \begin{tabular}{@{} l l l l l @{}}
    \toprule
    Component & \multicolumn{2}{l}{OD} & \multicolumn{2}{l}{Transmission} \\
    \midrule

    \textbf{Two ND Filters} 
        & 1   &   & 10\%   &   \\
        & 4   &   & 0.01\% &   \\

    \midrule
    \textbf{Five-position Filter Wheel} 
        & 0.2 &   & 63.1\% &   \\
        & 0.4 &   & 39.8\% &   \\
        & 0.6 &   & 25.1\% &   \\
        & 0.9 &   & 12.6\% &   \\
        & \multicolumn{2}{l}{Empty} & \multicolumn{2}{l}{1} \\

    \midrule
    \textbf{Eight-position Filter Wheel} 
        & 1   &   & 10\%   &   \\
        & 2   &   & 1\%    &   \\
        & 3   &   & 0.1\%  &   \\
        & 4   &   & 0.01\% &   \\
        & \multicolumn{2}{l}{Empty} & \multicolumn{2}{l}{1} \\
        & \multicolumn{2}{l}{Slots 6--8: Black} & \multicolumn{2}{l}{0} \\

    \bottomrule
  \end{tabular}

  \caption{Optical densities (OD) of all neutral density (ND) filters used in the setup.
  By combining these filters, the total attenuation can span a wide range, up to 
  $7.94\times10^{9}$. Note that "black" positions correspond to nearly zero transmission.}
   \label{tab:ND}
\end{table}

%% file: ref.bib
@article{trident2024sipm,
    author = "Zhi, Wei and Cao, Ruike and Tang, Jiannan and Wang, Mingxin and Tan, Yongqi and Wu, Weihao and Xu, Donglian",
    title = "{Front-end electronics development of large-area SiPM arrays for high-precision single-photon time measurement}",
    eprint = "2403.02948",
    archivePrefix = "arXiv",
    primaryClass = "physics.ins-det",
    doi = "10.1088/1748-0221/19/06/P06011",
    journal = "JINST",
    volume = "19",
    number = "06",
    pages = "P06011",
    year = "2024"
}

@article{trident2023hdom,
    author = "Zhi, Wei and Zheng, Jie and Tian, Wei and Xu, Donglian and Xiang, Xin",
    title = "{Preliminary Design of the Hybrid Digital Optical Module for TRIDENT}",
    doi = "10.22323/1.444.1213",
    journal = "PoS",
    volume = "ICRC2023",
    pages = "1213",
    year = "2023"
}

@misc{yyong2025,
  author       = {Zhang, Guangping and Yang, Yong},
  title        = {Design and Performance of the hDOM Motherboard for {TRIDENT Phase-I}},
  howpublished = {Poster contribution at the 39th International Cosmic Ray Conference (ICRC2025), Geneva, Switzerland. Available at: \url{https://indico.cern.ch/event/1258933/contributions/6485942/}},
  year         = {2025}
}

@article{Shao2025_TRIDENT,
  author        = {Shao, Hengbin and others},
  title         = {A Cost Effective Optimization of the hybrid-DOM Design for TRIDENT},
  eprint        = {2507.10256},
  archivePrefix = {arXiv},
  primaryClass  = {hep-ex},
  year          = {2025},
  month         = {7}
}

@article{Aguilar2011_ANTARES,
  author    = {J. A. Aguilar et al.},
  title     = {Time calibration of the ANTARES neutrino telescope},
  journal   = {Astropart. Phys.},
  volume    = {34},
  pages     = {539--549},
  year      = {2011},
  doi       = {10.1016/j.astropartphys.2010.10.004}
}

@misc{VITROVEX17,
  title        = {{VITROVEX} 17-inch glass spheres (technical data sheet)},
  author       = {Nautilus Marine Service GmbH},
  year         = {2017},
  url          = {https://www.nautilus-gmbh.com/VITROVEX},
  note         = {Accessed 2025-10-31}
}

@article{CockcroftWalton1932,
  author    = {J. D. Cockcroft and E. T. S. Walton},
  title     = {Experiments with High Velocity Positive Ions. {I}. Further Developments in the Method of Obtaining High Velocity Positive Ions},
  journal   = {Proceedings of the Royal Society A},
  volume    = {136},
  number    = {830},
  pages     = {619--630},
  year      = {1932},
  doi       = {10.1098/rspa.1932.0133}
}

@article{KM3NeT_DOM_2018,
  author  = {Leonora, E. and others},
  title   = {The Digital Optical Module of {KM3NeT}},
  journal = {Journal of Physics: Conference Series},
  volume  = {1056},
  number  = {1},
  pages   = {012031},
  year    = {2018},
  doi     = {10.1088/1742-6596/1056/1/012031},
  url     = {https://iopscience.iop.org/article/10.1088/1742-6596/1056/1/012031},
  note    = {KM3NeT Collaboration}
}

@article{IceCube_DOM_2006,
  author  = {Hanson, K. and Tarasova, O.},
  title   = {Design and production of the {IceCube} digital optical module},
  journal = {Nucl. Instrum. Meth. A},
  volume  = {567},
  number  = {1},
  pages   = {214--217},
  year    = {2006},
  doi     = {10.1016/j.nima.2006.05.091},
  note    = {\href{https://doi.org/10.1016/j.nima.2006.05.091}{DOI: 10.1016/j.nima.2006.05.091}, IceCube Collaboration}
}

@article{TRIDENT:2022hql,
    author = "Ye, Z. P. and others",
    collaboration = "TRIDENT",
    title = "{A multi-cubic-kilometre neutrino telescope in the western Pacific Ocean}",
    eprint = "2207.04519",
    archivePrefix = "arXiv",
    primaryClass = "astro-ph.HE",
    doi = "10.1038/s41550-023-02087-6",
    journal = "Nature Astron.",
    volume = "7",
    number = "12",
    pages = "1497--1505",
    year = "2023"
}

@book{Gaisser2016,
  author    = {Gaisser, T. K. and Engel, R. and Resconi, E.},
  title     = {Cosmic Rays and Particle Physics},
  edition   = {2nd},
  publisher = {Cambridge University Press},
  year      = {2016}
}

@article{DEAP:2017fgw,
    author = "Amaudruz, P. -A. and others",
    collaboration = "DEAP",
    title = "{In-situ characterization of the Hamamatsu R5912-HQE photomultiplier tubes used in the DEAP-3600 experiment}",
    eprint = "1705.10183",
    archivePrefix = "arXiv",
    primaryClass = "physics.ins-det",
    doi = "10.1016/j.nima.2018.12.058",
    journal = "Nucl. Instrum. Meth. A",
    volume = "922",
    pages = "373--384",
    year = "2019"
}

@article{Halzen:2010yj,
    author = "Halzen, Francis and Klein, Spencer R.",
    title = "{IceCube: An Instrument for Neutrino Astronomy}",
    eprint = "1007.1247",
    archivePrefix = "arXiv",
    primaryClass = "astro-ph.HE",
    doi = "10.1063/1.3480478",
    journal = "Rev. Sci. Instrum.",
    volume = "81",
    pages = "081101",
    year = "2010"
}

@article{Adrian-Martinez2016,
  author  = {S. Adri{\'a}n-Mart{\'\i}nez and others},
  title   = {Letter of intent for {KM3NeT} 2.0},
  journal = {J. Phys. G},
  volume  = {43},
  pages   = {084001},
  year    = {2016},
  doi     = {10.1088/0954-3899/43/8/084001},
  note    = {{KM3NeT Collaboration}}
}

@article{PRESCOTT1966173,
title = {A statistical model for photomultiplier single-electron statistics},
journal = {Nuclear Instruments and Methods},
volume = {39},
number = {1},
pages = {173-179},
year = {1966},
issn = {0029-554X},
doi = {https://doi.org/10.1016/0029-554X(66)90059-0},
url = {https://www.sciencedirect.com/science/article/pii/0029554X66900590},
author = {J.R. Prescott}
}

@article{Kalousis:2020PMTCalib,
  author         = {Kalousis, L. N. and de Andr{\'e}, J. P. A. M. and
                    Baussan, E. and Dracos, M.},
  title          = {A fast numerical method for photomultiplier tube calibration},
  journal        = {JINST},
  volume         = {15},
  year           = {2020},
  pages          = {P03023},
  doi            = {10.1088/1748-0221/15/03/P03023},
  eprint         = {1911.06220},
  archivePrefix  = {arXiv},
  primaryClass   = {physics.ins-det}
}

@article{Akashi-Ronquest:2019mlk,
    author = "Akashi-Ronquest, Michael and others",
    title = "{Triplet Lifetime in Gaseous Argon}",
    eprint = "1903.06706",
    archivePrefix = "arXiv",
    primaryClass = "physics.ins-det",
    doi = "10.1140/epja/i2019-12867-2",
    journal = "Eur. Phys. J. A",
    volume = "55",
    number = "10",
    pages = "176",
    year = "2019"
}

@article{Zhang:2025vvf,author  = {Zhang, Guangping and Yang, Yong and Xu, Donglian},
  title = {A waveform and time digitization mainboard prototype for the hybrid digital optical module of {TRIDENT} neutrino experiment},
  journal = {Radiation Detection Technology and Methods},
  year    = {2025},
  doi     = {10.1007/s41605-025-00619-4},
  url     = {https://link.springer.com/article/10.1007/s41605-025-00619-4},
  note    = {Published online: 19 November 2025}
}

@article{Timmer_2010,
  author  = {P. Timmer and E. Heine and H. Peek},
  title   = {Very low power, high voltage base for a photomultiplier tube for the {KM3NeT} deep sea neutrino telescope},
  journal = {JINST},
  volume  = {5},
  pages   = {C12049},
  year    = {2010},
  doi     = {10.1088/1748-0221/5/12/C12049}
}

@inproceedings{IceCube:2021_mDOM,
  author       = "{The IceCube Collaboration}",
  title        = "{Design and performance of the multi-PMT optical module for IceCube Upgrade}",
  booktitle    = "Proceedings of the 37th International Cosmic Ray Conference (ICRC 2021)",
  year         = "2021",
  volume       = "395",
  pages        = "1070",
  series       = "PoS",
  number       = "ICRC2021",
  publisher    = "Sissa Medialab",
  location     = "Berlin, Germany",
  doi          = "10.22323/1.395.1070"
}
